\documentclass{llncs}

\usepackage[utf8]{inputenc}
\usepackage[T1]{fontenc}
\usepackage[final]{graphicx}
\usepackage[labelsep=period]{caption}
\usepackage[hyphens]{url}
\usepackage[english]{babel}
\usepackage{amsmath}
\usepackage{algorithm}
\usepackage{algorithmic}
\usepackage{multicol}
\usepackage{color, colortbl}

\definecolor{Gray}{gray}{0.9}

\begin{document}

\title{Can FCA-based Recommender System Suggest a Proper Classifier?}

\author{
Yury Kashnitsky \and Dmitry I. Ignatov
}

\institute{
National Research University Higher School of Economics \\
Scientific-Educational Laboratory for Intelligent Systems and Structural Analysis \\ Moscow, Russia \\
\email{
ykashnitsky@hse.ru, dignatov@hse.ru}
}
\maketitle

\begin{abstract}
The paper briefly introduces multiple classifier systems and describes a new algorithm, which improves classification accuracy by means of recommendation of a proper algorithm to an object classification. This recommendation is done assuming that a classifier is likely to predict the label of the object correctly if it has correctly classified its neighbors. The process of assigning a classifier to each object is based on Formal Concept Analysis. 
We explain the idea of the algorithm with a toy example and describe our first experiments with real-world datasets.

\end{abstract}

\section{Introduction}
\label{intro}

\hspace{0.35cm} The topic of Multiple Classifier Systems (MCSs) is well studied in machine learning community \cite{izenman}. Such algorithms appear with different names -- mixture of experts, committee machines, classifier ensembles, classifier fusion and others. 

The underlying idea of all these systems is to train several (base) classifiers on a training set and to combine their predictions in order to classify objects from a test set \cite{izenman}. This idea probably dates back to as early as the 18$^{th}$ century. The Condorcet's jury theorem, that was formulated in 1785 in \cite{condorcet}, claims that if a population makes a group decision and each voter most likely votes correctly, then adding more voters increases the probability that the majority decision is correct. The probability that the majority votes correctly tends to 1 as the number of voters increases. Similarly, if we have multiple weak classifiers (meaning that classifier's error on its training data is less than 50\% but greater than 0\%), we can combine their predictions and boost the classification accuracy as compared to those of each single base classifier. 

Among the most popular MCSs are bagging \cite{bagging}, boosting \cite{adaboost}, random forests \cite{rand_forests}, and stacked generalization (or stacking) \cite{stacking}. 

In this paper, we present one more algorithm of such type -- Recommender-based Multiple Classifier System (RMCS). Here the underlying proposition is that a classifier is likely to predict the label of the object from a test set correctly if it has correctly classified its neighbors from a training set. 

The paper is organized as follows. In chapter \ref{sec:MCS}, we discuss bagging, boosting and stacking. In Section \ref{sec:FCA}, we introduce basic definitions of Formal Concept Analysis (FCA). Section \ref{sec:example} provides an example of execution of the proposed RMCS algorithm on a toy synthetic dataset. Then, Section \ref{sec:RMCS} describes the RMCS algorithm itself. Further, the results of the experiments with real data are presented. Section \ref{sec:conclusion} concludes the paper.


\section{Multiple Classifier Systems}
\label{sec:MCS}

In this chapter, we consider several well-known multiple classier systems. 
\subsection{Bagging}
\hspace{0.35cm} The \textit{bootstrap sampling} technique has been used in statistics for many years. \textit{Bootstrap aggregating}, or \textit{bagging}, is one of the applications of bootstrap sampling in machine learning. As sufficiently large data sets are often expensive or impossible to obtain, with bootstrap sampling, multiple random samples are created from the source data by sampling with replacement. Samples may overlap
or contain duplicate items, yet the combined results are usually more accurate than a single sampling of the entire source data achieves.

In machine learning the bootstrap samples are often used to train classifiers.
Each of these classifiers can classify new instances making a prediction; then predictions are combined to  obtain a final classification.

The aggregation step of bagging is only helpful if the classifiers are different. This only happens if small changes in the training data can result in large changes in the resulting classifier -- that is, if the learning method is unstable \cite{bagging}.

\subsection{Boosting}
\hspace{0.35cm} The idea of \textit{boosting} is to iteratively train classifiers with a weak learner (the one with error better than 50\% but worse than 0\%) \cite{boosting1}. After each classifier is trained, its accuracy is measured, and misclassified instances are emphasized. Then the algorithm trains a new classifier on the modified dataset. At classification time, the boosting classifier combines the results from the individual classifiers it trained.

Boosting was originally proposed by Schapire and Freund \cite{boosting2,boosting3}. In their \textit{Adaptive Boosting}, or \textit{AdaBoost}, algorithm, each of the training instances starts with a weight that tells the base classifier its relative importance \cite{adaboost}. At the initial step the weights of $n$  instances are evenly distributed as $\frac{1}{n}$
The individual classifier training algorithm should take into account these weights, resulting in different classifiers after each round of reweighting and reclassification. Each classifier also receives a weight based on its accuracy; its output at classification time is multiplied by this weight.

Freund and Schapire proved that, if the base classifier used by AdaBoost has an error rate of just slightly less than 50\%, the training error of the meta-classifier will approach zero exponentially fast \cite{adaboost}. For a two-class problem the base classifier only needs to be slightly better than chance to achieve this error rate. For problems with more than two classes less than 50\% error is harder to achieve. Boosting appears to be vulnerable to overfitting. However, in tests it rarely overfits excessively \cite{dietterich}.

\subsection{Stacked generalization}
\hspace{0.35cm} In \textit{stacked generalization}, or \textit{stacking},  each individual classifier is called a \textit{level-0 model}. Each may vote, or may have its output sent to a \textit{level-1 model} -- another classifier that tries to learn which level-0 models are most reliable. Level-1 models are usually more accurate than simple voting, provided they are given the class probability distributions from the level-0 models and not just the single predicted class \cite{stacking}.


\section{Introduction to Formal Concept Analysis}
\label{sec:FCA}

\subsection{Main definitions}

\hspace{0.35cm} A \textit{formal context} in FCA is a triple $K = (G, M, I)$, where $G$ is a set of objects, $M$ is a set of attributes, and the binary relation $I \subseteq G \times M$ shows which object possesses which attribute. $gIm$ denotes that object $g$ has attribute $m$. For subsets of objects and attributes $A \subseteq G$ and $B \subseteq M$ \textit{Galois operators} are defined as follows:
\begin{equation*} 
\begin{split}
A' = \{m \in M\ |\ gIm\ \forall g \in A \}, \\
B' = \{g \in G\ |\ gIm\ \forall m \in B\}. 
\end{split}
\end{equation*}

A pair $(A, B)$ such that $A \subseteq G, B \subseteq M , A' = B$ and $B' = A$, is called a \textit{formal concept} of a context $K$. The sets $A$ and $B$ are closed and called the \textit{extent} and the \textit{intent} of a formal concept $(A, B)$ respectively. For the set of objects $A$ the set of their common attributes $A'$ describes the similarity of objects of the set $A$ and the closed set $A''$ is a cluster of similar objects (with the set of common attributes $A’$) \cite{FCA}.
 
The number of formal concepts of a context $K = (G, M, I)$ can be quite large ($2^{min\{|G|, |M|\}}$ in the worst case), and the problem of computing this number is \#P-complete \cite{hard}. There exist some ways to reduce the number of formal concepts, for instance, choosing concepts by stability, index or extent size \cite{concept_reduce}.

For a context $(G, M, I)$, a concept $X = (A, B)$ is \textit{less general than or equal to} a concept $Y = (C, D)$ (or $X \leq Y$) if $A \subseteq
C$ or, equivalently, $D \subseteq B$. For two concepts $X$ and $Y$ such that $X \leq Y$ and there is no concept $Z$ with $Z \neq X, Z \neq Y, X \leq Z \leq Y$, the concept $X$ is called a \textit{lower neighbor} of $Y$,
and $Y$ is called an \textit{upper neighbor} of $X$. This relationship is denoted by $X \prec Y$. Formal concepts, ordered by this relationship, form a \textit{complete concept lattice} which might be represented by a \textit{Hasse diagram} \cite{lattices}. Several algorithms for building formal concepts (including $Close\ by\ One$) and constructing concept lattices are studied also in \cite{lattices}.

One can address to \cite{FCA} and \cite{iceberg} to find some examples of formal contexts, concepts and lattices with their applications. Chapter \ref{sec:example} also shows the usage of FCA apparatus in a concrete task.

However, in some applications there is no need to find all formal concepts of a formal context or to build the whole concept lattice. Concept lattices, restricted to include only concepts with frequent intents, are called $iceberg\ lattices$. They were shown to serve as a condensed representation of association rules and frequent itemsets in data mining \cite{iceberg}.  

Here we modified the $Close\ by\ One$ algorithm slightly in order to obtain only the upper-most concept of a formal context and its lower neighbors. The description of the algorithm and details of its modification is beyond the scope of this paper. 


\section{A toy example}
\label{sec:example}
\hspace{0.35cm} Let us demonstrate the way RMCS works with a toy synthetic dataset shown in Table \ref{toy_dataset}. We consider a binary classification problem with 8 objects comprising a training set and 2 objects in a test set. Each object has 4 binary attributes and a target attribute (class). 
Suppose we train 4 classifiers on this data and try to predict labels for objects 9 and 10. 
 
Using FCA terms, we denote by $G = \{1,2,3,4,5,6,7,8,9,10\}$ --- the whole set of objects, $G_{test} = \{9,10\}$ --- the test set, $G_{train} = G \textbackslash G_{test}$  --- the training set, $M = \{m_1,m_2,m_3,m_4\}$ --- the attribute set, $C = \{cl_1, cl_2, cl_3, cl_4\}$ --- the set of classifiers. 

\begin{table}
	\begin{minipage}[b]{0.45\linewidth}
		\caption{A sample data set of 10 objects with 4 attributes and 1 			binary target class}
		\label{toy_dataset}
		\begin{tabular}{|c|c|c|c|c|c|}
		
		\hline
		G/M& $m_1$ & $m_2$ & $m_3$ & $m_4$ & Label\\\hline
		1 & $\times$ & $\times$ &   & $\times$ & 1 \\\hline
		2 & $\times$ &   &   & $\times$ & 1 \\\hline
		3 &   & $\times$ & $\times$ &   & 0 \\\hline
		4 & $\times$ &   & $\times$ & $\times$ & 1 \\\hline
		5 & $\times$ & $\times$ & $\times$ &   & 1 \\\hline
		6 &   & $\times$ & $\times$ & $\times$ & 0 \\\hline
		7 & $\times$ & $\times$ & $\times$ &   & 1 \\\hline
		8 &   &   & $\times$ & $\times$ & 0 \\\hline
		9 & $\times$ & $\times$ & $\times$ & $\times$ & ? \\\hline
		10&   & $\times$ &   & $\times$ & ? \\\hline
		\end{tabular}
		\end{minipage}
		\begin{minipage}[b]{0.45\linewidth}
		\caption{A classification context}
		\label{class_context}
		\begin{tabular}{|c|c|c|c|c|}
		\hline
		G/C& $cl_1$ & $cl_2$ & $cl_3$ & $cl_4$ \\\hline
		1 & $\times$ &  &  $\times$ & $\times$ \\\hline
		2 &  & $\times$   & $\times$ & \\\hline
		3 & $\times$  &  & & $\times$   \\\hline
		4 &  & $\times$   & $\times$ & \\\hline
		5 & $\times$ & $\times$ &  &   \\\hline
		6 & $\times$  & $\times$ &  & $\times$ \\\hline
		7 &  & $\times$ & & $\times$ \\\hline
		8 &   &  $\times$ & $\times$ & $\times$  \\\hline
		\end{tabular}
		\end{minipage}
\end{table}

Here we run leave-one-out cross-validation on this training set for 4 classifiers. Further, we fill in Table \ref{class_context}, where a cross for object $i$ and classifier $cl_j$ means that $cl_j$ correctly classifies object $i$ in the process of cross-validation. To clarify, a cross for object 3 and classifier $cl_4$ means that after being trained on the whole training set but object 3 (i.e. on objects $\{1,2,4,5,6,7,8\}$), classifier $cl_4$ correctly predicted the label of object 3.

\begin{figure}[ht]
\centering
\includegraphics[width=\linewidth]{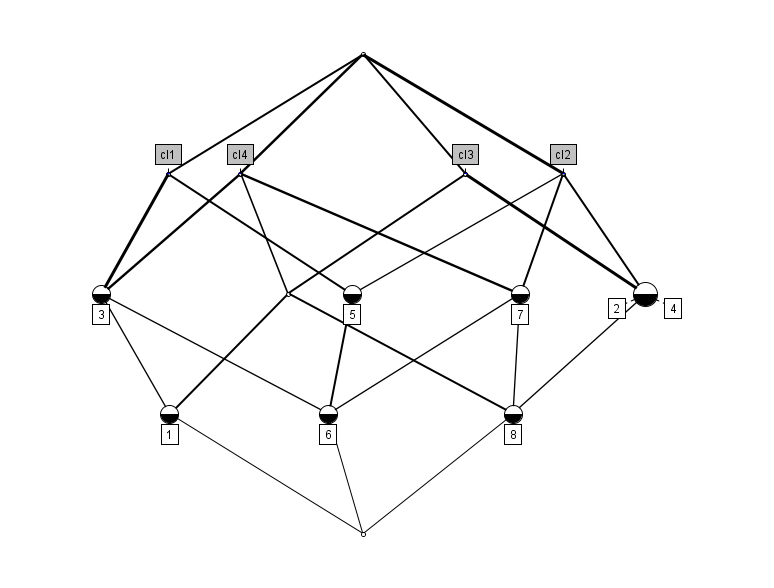}
\caption{The concept lattice of the classification context}
\label{fig:lattice}
\end{figure}

Let us consider Table \ref{class_context} as a formal context with objects $G$ and attributes $C$ (so now classifiers play the role of attributes). We refer to it as classification context. The concept lattice for this context is presented in Fig. \ref{fig:lattice}.

As it was mentioned, the number of formal concepts of a context $K =$ $(G, M, I)$ can be exponential in the worst case. But for the toy example it is possible to draw the whole lattice diagram. Thankfully, we do not need to build the whole lattice in RMCS algorithm --- we only keep track of its top concepts. 

Here are these top concepts: $(G,\emptyset)$, $(\{1,3,5,6\},\{cl_1\})$, $(\{2,4,5,6,7,8\}, \{cl_2\})$, $(\{1,2,4,8\},\{cl_3\})$, $(\{1,3,6,7,8\}, \{cl_4\})$.

To classify objects from $G_{test}$, we first find their $k$ nearest neighbors from $G_{train}$ according to some distance metric. In this case, we use $k=3$ and Hamming distance. In these conditions, we find that three nearest neighbors of object 9 are 4, 5 and 7, while those of object 10 are 1, 6 and 8. 

Then, we take these sets of nearest neighbors $Neighb_9 = \{4,5,7\}$ and \\ $Nieghb_{10} = \{1,6,8\}$, and find maximal intersections of these sets with the extents of formal concepts presented above (ignoring the concept $(G,\emptyset)$). The intents (i.e. classifiers) of the corresponding concepts are given as recommendations for the objects from $G_{test}$. The procedure is summarized in Table \ref{recommend_clf}.

\begin{center}
	\begin{table}
		\caption{Recommending classifiers for objects from $G_{test}$}
		\label{recommend_clf}
		\begin{tabular}{|c|p{1.3cm}|p{1.3cm}|p{1.3cm}|p{1.5cm}|p{3.1cm}|p{2.4cm}|}
		\hline
		$G_{test}$& $1^{st}$ \par nearest neighbor & $2^{nd}$ \par nearest neighbor & $3^{rd}$ \par nearest neighbor & Neighbors & Classification concept which extent gives the maximal intersection with the Neighbors set & Recommended \par classifier\\\hline
		9 & 4 & 5 &  7 & $\{4,5,7\}$ & $(\{2,4,5,6,7,8\}, \{cl_2\})$ & $cl_2$ \\\hline
		10 & 1 & 6 & 8 & $\{1,6,8\}$ & $(\{1,3,6,7,8\}, \{cl_4\})$ & $cl_4$\\\hline
		\end{tabular}	
		\end{table}
\end{center}

Finally, the RMCS algorithm predicts the same labels for objects 9 and 10 as classifiers $cl_2$ and $cl_4$ do correspondingly.

Lastly, let us make the following remarks:

\begin{enumerate}
\item We would not have ignored the upper-most concept with extent $G$ if it did not have an empty intent. That is, if we had the top concept of the classification context in a form $(G, \{cl_j\})$ it would mean that $cl_j$ correctly classified all objects from the training set and we would therefore recommend it to the objects from the test set. 

\item One more situation might occur that two or more classifiers turn out to be equally good at classifying objects from $G_{train}$. That would mean that the corresponding columns in classification table are identical and, therefore, the intent of some classification concept is comprised of more than one classifier. In such case, we do not have any argument for preferring one classifier to another and, hence, the final label would be defined as a result of voting procedure among the predicted labels of these classifiers. 

\item Here we considered an input dataset with binary attributes and a binary target class. However, the idea of the RMCS algorithm is still applicable for datasets with numeric attributes and multi-class classification problems. 
\end{enumerate}  

\section{Recommender-based Multiple Classifier System}
\label{sec:RMCS}

\hspace{0.35cm} In this section, we discuss the Recommender-based Multiple Classifier System (RMCS). The pseudocode of the RMCS algorithm is presented in the listing Algorithm  \ref{alg:RMCS}.

The inputs for the algorithm are the following:
\begin{enumerate}
\item $\{X_{train}, y_{train}\}$ --- is a training set, $X_{test}$ --- is a test set;
\item $C = \{cl_1, cl_2,..., cl_K\}$ --- is a set of $K$ base classifiers. The algorithm is intended to perform a classification accuracy exceeding those of base classifiers;
\item $dist(x_1, x_2)$ --- is a distance function for objects which is defined in the attribute space. This might be the Minkowski (including Hamming and Euclidean) distance, the distance weighted by attribute importance and others.
\item $k, n\_fold$ --- are parameters. Their meaning is explained below;
\item $topCbO(context)$ --- is a function for building the upper-most concept of a formal context and its lower neighbors. Actually, it is not an input for the algorithm but RMCS uses it.
\end{enumerate}
The algorithm includes the following steps:

\begin{enumerate}
\item Cross-validation on the training set. All $K$ classifiers are trained on $n\_folds-1$ folds of $X_{train}$. Then a classification table (or context) is formed where a cross is put for object $i$ and classifier $cl_j$ if $cl_j$ correctly classifies object $i$ after training on $n\_folds-1$ folds (where object $i$ belongs to the rest fold);
\item Running base classifiers. All $K$ classifiers are trained on the whole $X_{train}$. Then, a table of predictions is formed where $(i,j)$ position keeps the predicted label for object $i$ from $X_{test}$ by classifier $cl_j$;
\item Building top formal concepts of the classification context. The $topCbO$ algorithm is run in order to build upper formal concepts of a classification context. These concepts have the largest possible number of objects in extents and minimal possible number of classifiers in their intents (not counting the upper-most concept);
\item Finding neighbors of the objects from $X_{test}$. The objects from the test set are processed one by one.
For every object from $X_{test}$ we find its $k$ nearest neighbors from $X_{train}$ according to the selected metric $sim(x_1, x_2)$. Let us say these $k$ objects form a set $Neighbors$. Then, we search for a concept of a classification context which extent yields maximal intersection with the set $Neighbors$. If the intent of the upper-most concept is an empty set (i.e., no classifier correctly predicted the labels of all objects from $X_{train}$, which is mostly the case), then the upper-most concept $(G,\emptyset)$ is ignored. Thus, we select a classification concept, and its intent is a set of classifiers $C_{sel}$;
\item Classification. If $C_{sel}$ consists of just one classifier, we predict the same label for the current object from $X_{test}$ as this classifier does. If there are several selected classifiers, then the predicted label is defined by majority rule.
\end{enumerate}

\begin{algorithm}
\textbf{Input:} $\{X_{train}, y_{train}\}, X_{test}$ --- are training and test sets,
$C = \{cl_1, cl_2,..., cl_K\}$ --- is a set of base classifiers, $topCbO(context, n)$ --- is a function for building the upper-most concept of a formal context and its lower neighbors, $dist(x_1, x_2)$ --- is a distance function defined in the attribute space, 
$k$ --- is a parameter (the number of neighbors), $n\_fold$ --- is the number of folds for cross-validation on a training set\\
\textbf{Output:} $y_{test}$ --- are predicted labels for objects from $X_{test}$
\begin{algorithmic}
\STATE $train\_class\_context = [\ ][\ ]$ --- is a 2-D array
\STATE $test\_class\_context = [\ ][\ ]$ --- is a 2-D array
\FOR{$i \in 0 \dots len(X_{train})-1$}
\FOR{$cl \in 0 \dots len(C)-1$}
\STATE train classifier $cl$ on $(n\_fold-1)$ folds not including object $X_{train}[i]$
\STATE $pred$ = predicted label for $X_{train}[i]$ by classifier $cl$  
\STATE $train\_class\_context[i][cl] = (pred == y_{train}[i])$ 
\ENDFOR
\ENDFOR

\FOR{$cl \in 0 \dots len(C)-1$}
\STATE train classifier $cl$ on the whole $X_{train}$
\STATE $pred$ = predicted labels for $X_{test}$ by classifier $cl$  
\STATE $test\_class\_context[:][cl] = pred$ 
\ENDFOR

\STATE $top\_concepts = topCbO(class\_context)$  
\FOR{$i \in 0 \dots len(X_{test})-1$}
\STATE $Neighbors = k$ nearest neighbors of $X_{test}[i]$ from $X_{train}$ according to $sim(x_1,x_2)$
\STATE $concept = argmax(c.extent \  \cap \ Neighbors), c \in top\_concepts$
\STATE $C_{sel} = concept.intent$
\STATE $labels = $ predictions for $X_{test}[i]$ made by classifiers from $C_{sel}$
\STATE $y_{test}[i] = argmax(count\_freq(labels))$
\ENDFOR
\end{algorithmic}
\caption{Recommender-based Multiple Classifier System}
\label{alg:RMCS}
\end{algorithm}


\section{Experiments}
\hspace{0.35cm} The algorithm, described above, was implemented in Python 2.7.3 and tested on a 2-processor machine (Core i3-370M, 2.4 HGz) with 3.87 GB RAM.

We used four UCI datasets in these experiments - \verb"mushrooms", \verb"ionosphere", \verb"digits", and \verb"nursery".\footnote{\url{http://archive.ics.uci.edu/ml/datasets}} Each of the datasets was divided into training and test sets in proportion 70:30.
 
\begin{table}
		\caption{Classification accuracy of 6 algorithms on 4 UCI datasets: mushrooms (1), ionosphere (2), digits (3), and nursery (4)}
		\label{table_tests}
		\begin{tabular}{|c|p{2.3cm}|p{2cm}|p{2.3cm}|>{\columncolor{Gray}}p{1.8cm}|p{2.1cm}|p{2cm}|}
		\hline
		Data & SVM, \par RBF kernel  \par (C=1, $\gamma$=0.02) & Logit \par (C=10) &  kNN \par (euclidean, \par k=3)&  RMCS \par (k=3, \par n\_folds=4) & Bagging SVM \par (C=1, $\gamma$=0.02) \par 50 estimators & AdaBoost \par on decision \par stumps, \par 50 iterations\\\hline
		1 & 0.998 \par t=0.24 sec.  & 0.996  \par t=0.17 sec. & 0.989 \par t=1.2*$10^{-2}$ sec. & 0.997 \par t=29.45 sec. & 0.998 \par t=3.35 sec. & 0.998 \par t=44.86 sec. \\\hline
		2 & 0.906 \par t=5.7*$10^{-3}$ sec. & 0.868 \par t=$10^{-2}$ sec. & 0.858 \par t=8*$10^{-4}$ sec. & 0.933 \par t=3.63 sec. & 0.896 \par t=0.24 sec. & 0.934 \par t=22.78 sec.\\\hline
		3 & 0.917  \par t=0.25 sec. & 0.87 \par t=0.6 sec.& 0.857 \par t=1.1*$10^{-2}$ sec.& 0.947 \par t=34.7 sec.& 0.92 \par t=4.12 sec. & 0.889 \par t=120.34 sec.\\\hline
		4 & 0.914 \par t=3.23 sec. & 0.766 \par t=0.3 sec.& 0.893 \par t=3.1*$10^{-2}$ sec.& 0.927 \par t=220.6 sec.& 0.913 \par t=38.52 sec. & 0.903 \par t=1140 sec.\\\hline
		
		\end{tabular}	
\end{table}

\begin{table}
		\begin{tabular}{|c|p{2.4cm}|p{2cm}|p{2.25cm}|>{\columncolor{Gray}}p{1.8cm}|p{2.1cm}|p{2cm}|}
		\hline
		Data & SVM, \par RBF kernel  \par (C=$10^3$, $\gamma$=0.02) & Logit \par (C=$10^3$) &  kNN \par (minkowski, p=1, k=5)&  RMCS \par (k=5, \par n\_folds=10) & Bagging SVM \par (C=$10^3$, $\gamma$=0.02) \par 50 estimators & AdaBoost \par on decision \par stumps, \par 100 iterations\\\hline
		1 & 0.998 \par t=0.16 sec.  & 0.999  \par t=0.17 sec. & 0.999 t=1.2*$10^{-2}$sec. & 0.999 \par t=29.45 sec. & 0.999 \par t=3.54 sec. & 0.998 \par t=49.56 sec. \\\hline
		2 & 0.906 \par t=4.3*$10^{-3}$ sec. & 0.868 \par t=$10^{-2}$ sec. & 0.887 \par t=8*$10^{-4}$ sec. & 0.9 \par t=3.63 sec. & 0.925 \par t=0.23 sec. & 0.934 \par t=31.97 sec.\\\hline
		3 & 0.937  \par t=0.22 sec. & 0.87 \par t=0.6 sec.& 0.847 \par t=1.1*$10^{-2}$ sec.& 0.951 \par t=34.7 sec.& 0.927 \par t=4.67 sec. & 0.921 \par t=131.6 sec.\\\hline
		4 & 0.969 \par t=2.4 sec. & 0.794 \par t=0.3 sec.& 0.945 \par t=3*$10^{-2}$ sec.& 0.973 \par t=580.2 sec. & 0.92 \par t=85.17 sec. & 0.912 \par t=2484 sec.\\\hline
		
		\end{tabular}	
\end{table}

We ran 3 classifiers implemented in \verb"SCIKIT-LEARN" library \footnote{http://scikit-learn.org}(written in Python) which served as base classifiers for the RMCS algorithm as well. These were a Support Vector Machine with Gaussian kernel (\verb"svm.SVC()" in \verb"Scikit"), logistic regression (\verb"sklearn.linear_model.LogisticRegression()") and k Nearest Neighbors classifier (\verb"sklearn.neighbors.classification."\\\verb"KNeighborsClassifier()"). 

The classification accuracy of each classifier on each dataset is presented in Table \ref{table_tests} along with special settings of parameters. Moreover, for comparison, the results for \verb"Scikit"'s implementation of bagging with SVM as a base classifier and AdaBoost on decision stumps \footnote{https://github.com/pbharrin/machinelearninginaction/tree/master/Ch07} are presented. 

As we can see, RMCS outperformed its base classifiers in all cases, while it turned out to be better than bagging only in case of multi-class classification problems (datasets \verb"digits" and \verb"nursery").

\section{Conclusion}
\label{sec:conclusion}
\hspace{0.35cm} In this paper, we described the underlying idea of multiple classifier systems, discussed bagging, boosting and stacking. Then, we proposed a multiple classifier system which turned out to outperform its base classifiers and two particular implementations of bagging and AdaBoost in two multi-class classification problems. 

Our further work on the algorithm will continue in the following directions:
exploring the impact of different distance metrics (such as the one based on attribute importance or information gain) on the algorithm's performance, experimenting with various types of base classifiers, investigating the conditions preferable for RMCS (in particular, when it outperforms bagging and boosting), improving execution time of the algorithm and analyzing RMCS's overfitting.  

\newpage 
\subsubsection*{Acknowledgements.}
The authors would like to thank their colleague from Higher School of Economics, Sergei Kuznetsov, Jaume Baixeries and Konstantin Vorontsov for their inspirational discussions which directly or implicitly influenced this study.

\bibliographystyle{splncs}

\end{document}